\documentclass[12pt]{article}
\usepackage{graphicx}
\title{Active learning for seismic processing parameterisation, with an application to first break picking}
\author{Alan Richardson (Ausar Geophysical)}
\begin{document}
\maketitle
\begin{abstract}
Parameter values for seismic processing steps are often chosen on a regular grid of samples and interpolated. Active learning instead attempts to optimally select the samples on which parameter values are chosen. For parameters that do not vary smoothly, this often reduces the number of samples that need to be labelled in order to achieve a desired accuracy on the whole dataset. In regression tasks this is typically achieved using a query by committee strategy that selects the samples on which a committee of models is most uncertain. I implement such a strategy for the first break picking task, where the parameters to be chosen are the centre and width of the picking window for each trace. For the committee members I use the centre of the picking window and three popular picking algorithms. Applying this to a real dataset, and with samples corresponding to shot gathers, the active learning approach primarily selects gathers near a jump in the first breaks, and achieves similar levels of accuracy on the whole dataset with about half the number samples picked as when the samples are randomly selected.
\end{abstract}
\section{Introduction}

\subsection{Active learning}
Active learning refers to semi-supervised machine learning in the situation where an oracle (such as a human) can be queried to add new samples to the labelled portion of the dataset (also known as the training dataset). It uses the samples that have already been labelled and those that are yet to be labelled to choose the samples that it would be most helpful for the oracle to label in order to learn to correctly label the entire dataset \cite{settles2012active}. Previous studies show that active learning can reduce the number of labelled samples required to attain a target accuracy, but it is best suited to situations where the labels do not vary smoothly \cite{castro2005faster}.

As with other machine learning methods, the majority of research on active learning has concentrated on the application to classification tasks. When the task is instead regression, one approach dominates, which is to query the oracle on the samples where the value is most uncertain, although alternatives exist \cite{10.1007/978-3-319-46562-3_24}. One means of calculating this uncertainty, known as query by committee, fits several models (such as randomly initialised neural networks) to the labelled samples and then estimates the uncertainty on unlabelled samples as the spread in the values predicted by the models.

\subsection{Seismic processing}
Seismic imaging allows remote sensing of a material's mechanical properties by analysing the response to waves passed through it. A common application is determining underground geologic structure by moving an energy source across the surface above the target area, emitting a wave at discrete locations (shots), and recording the response at receiver locations, with the response at one receiver known as a trace and a collection of traces being called a gather. An essential component of this method is the processing of the recorded response. This is typically a time-consuming and labour-intensive process that consists of multiple steps to perform tasks such as noise attenuation. The parameter values for each step need to be chosen, and often vary spatially across the dataset. As data volumes are typically large, this is frequently accomplished by choosing the parameter values on a regular grid and interpolating between them. This can, however, lead to both wasted time, as the same parameter values might be appropriate for large stretches of the data, and poor results, as the grid may not be fine enough to capture abrupt changes.

Recently many attempts have been made to use machine learning to reduce the effort required to parameterise seismic processing steps. The most common approach is to perform processing steps using a deep neural network. The goal is typically to train a network that will perform the desired processing task (such as denoising) itself, but it may alternatively be that the network estimates appropriate parameters for a conventional algorithm designed for the task. The network is either pretrained on a large collection of other datasets, or is trained on a labelled portion of the dataset to which it will be applied. Pretraining has the advantage of being quick to apply, but there is a large risk that the model will not perform well on the target dataset, and, if the results are not satisfactory, there is usually no way of adjusting the model to improve them. Training on a portion of the target dataset provides greater confidence that the results will be acceptable, but shares the same concerns regarding labelling as traditional approaches if it is done on a regular grid (wasting effort when samples are similar to ones that have already been labelled, and missing abrupt changes). Another machine learning-based approach to parameterisation, described in \cite{bekara2019automatic} for the task of denoising, uses attributes extracted from the data, and a machine learning method, to identify locations where the parameters may not be appropriate. Thus could allow iteratively growing the training dataset to include locations where the parameters are not yet satisfactory, but \cite{bekara2019automatic} does not discuss a means of prioritising the locations to add to the training dataset.

\subsection{First break picking}
One of the most common processing steps, from large 3D marine surveys to small 2D engineering surveys on land, is to identify the first time at which the response wave arrives at each receiver, known as first break picking. It is also one of the most labour intensive, requiring detailed manual work on the entire dataset.

The first breaks will tend to occur later at receivers that are further from the source. They may appear to form a smooth surface, however a variable near surface and irregular source and receiver placement can cause shifts, especially on land, that are both difficult and important to capture. A typical workflow involves manually picking first breaks on a coarse regular grid. An interpolation between these picks provides a rough estimate of the pick time for all traces. Applying a picking algorithm, such as those described in \cite{sabbione2010automatic}, in a window around this estimate will often increase the accuracy by including the shifts but may pick incorrectly. This is followed by a detailed examination of these picks to manually correct mistakes.

\subsection{Hypothesis}

Seismic processing parameterisation closely matches the situation that active learning is designed to address. I therefore hypothesise that active learning can be used to reduce wasted time and improve accuracy by selecting samples for human operators to label with the correct parameters in a way that is more effective than the simple regular grid or random sampling that is often used currently. To test this hypothesis I will apply a query by committee active learning approach to first break picking.

\section{Method}
Using active learning to make first break picking more efficient could be achieved in multiple ways. The way that I propose has a similar workflow to conventional picking, but chooses which gather to pick next based on the disagreement between a committee of different picking algorithms rather than using a regular grid.

\begin{figure}
        \includegraphics{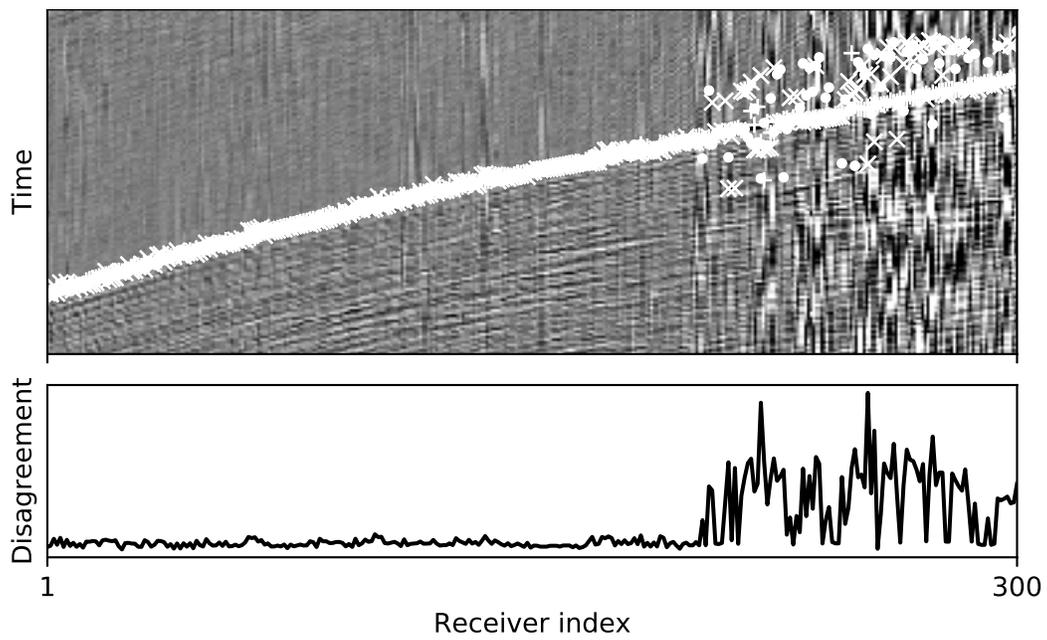}
        \caption{Picking algorithms on the committee respond differently to noise, causing there to be a large difference between the times picked by each (indicated by a different marker symbol for each algorithm).}
        \label{fig:noise_distance}
\end{figure}

Many picking algorithms have been proposed over the past few decades. Most of these could be used in the committee. I choose to use the three described in \cite{sabbione2010automatic}. These are the Modified Coppens, Entropy, and Fractal Dimension methods. The Coppens method assigns the first break pick to the time at which the recorded amplitude is largest compared to the amplitudes recorded before it. The other two methods assume that the entropy and fractal dimension, respectively, change when the recording transitions from noise to noise plus signal, and so identify the point at which this occurs as the first break. By relying on different assumptions, the different committee members will tend to respond differently to changes in the data, such as an increase in noise. This can be seen in Figure \ref{fig:noise_distance}. We can thus use disagreement among the committee to identify where human attention to correctly pick the first breaks is needed. The disagreement is measured as the sum of the absolute pairwise differences between the pick times of all of the committee members.

As is common in traditional picking workflows, I adjust the outputs of the picking algorithms to the nearest amplitude extremum (peak or trough). This improves consistency between traces and also among the picking algorithms, disregarding inconsequential differences between their picked times.

\begin{figure}
        \includegraphics{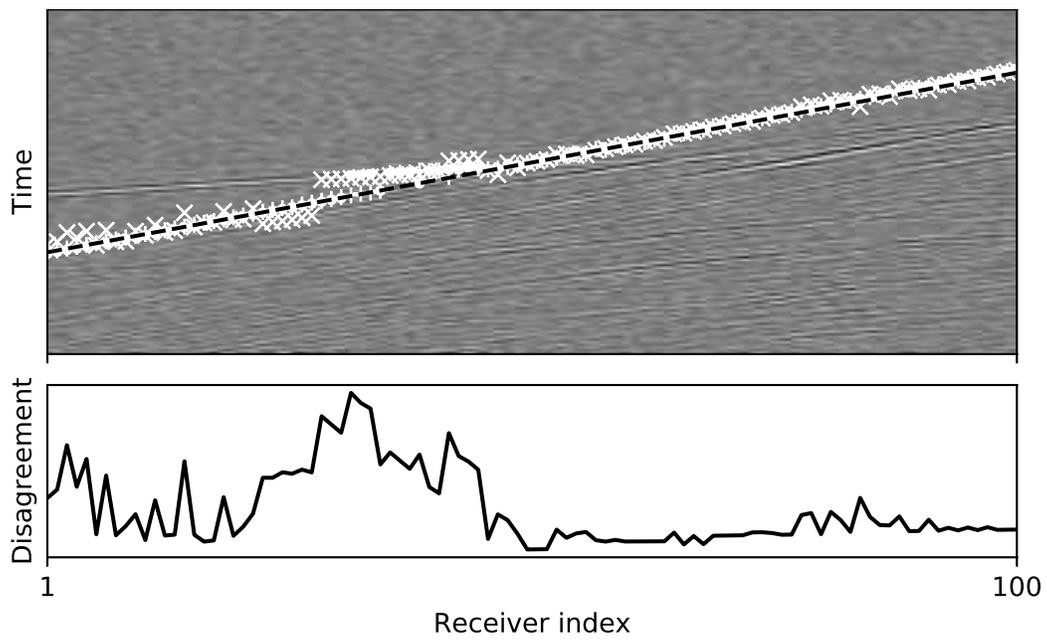}
        \caption{When the predicted centre (indicated by the dashed line) does not follow the actual first breaks, the disagreement between it and the other members of the committee (the picks of the picking algorithms) will increase.}
        \label{fig:centre_distance}
\end{figure}

If allowed to pick any time from a whole trace, picking algorithms often pick the wrong time for the first break, such as picking the second arrival or picking a burst of noise that occurs earlier in the trace. To mitigate this problem, traditional picking workflows and the proposed method only apply the picking algorithms in a window around a rough estimate of the pick time. The centre of the window is obtained by interpolating between picks that have already been made. I use piecewise linear interpolation. I add the predicted centre time to the committee as another estimate of the pick time. This causes the human operator to be asked to pick gathers where the predicted centre departs from the times picked by the picking algorithms. As shown in Figure \ref{fig:centre_distance}, this ensures that changes in the first break pick time are captured by the model that predicts the window centre. In my proposal I set the width of the window for each trace to be a fraction (determined by cross-validation) of the largest value that will still result in all of the picking algorithms in the committee choosing the correct pick time for that trace to within a tolerance that I set as the dominant period of the dataset. This window size is also linearly interpolated between picked traces so that the windows at unpicked traces can be calculated.

It may seem that the picking algorithms need to be applied within the windows on every trace in the dataset after each gather is picked, in order to choose the optimal gather to pick next. The computational cost of that may be substantial. While trading human time for extra computation is often worthwhile, I propose to reduce the cost by only considering a randomly selected sample of unpicked gathers as candidates for the next gather to pick. After each gather is picked I choose a new random selection of gathers for the committee to consider.

When picking has finished there is likely to still be some disagreement between the members of the committee on the pick time for each trace. I use the median of these values as the final pick.

The method thus uses query by committee active learning to determine the value of two parameters for each trace: the centre of the picking window and its length. It does this by optimising the selection of gathers on which a human operator picks the first break time for each trace. This pick time gives the centre parameter for that trace. It also gives the length parameter, as this is calculated from the largest window that would still give the correct pick time (within a tolerance) for all of the picking algorithms in the committee. Linear interpolation is used between picked traces to calculate the centre and window length for traces in unpicked gathers. The committee consists of three picking algorithms that are applied in the picking window, and the linearly interpolated centre. Picking continues until a termination condition is met, which might be that the highest disagreement is below a specified threshold, or that the number of picks has reached a specified maximum. For clarity, the steps of the algorithm are listed below.

\begin{enumerate}
\item{Randomly select one gather and obtain its picks from the oracle}
\item{Randomly select $N$ gathers from those that have not yet been picked}
\item{Linearly interpolate the centre and window length parameters from the picked traces to the gathers selected in 2}
\item{Apply the committee picking algorithms within the picking windows defined by the centre and window length on the gathers selected in 2}
\item{Calculate the disagreement for each trace in these gathers as the sum of the absolute pairwise differences between the picks by all of the committee members}
\item{Calculate the disagreement for each of the selected gathers by summing the disagreement of their traces}
\item{Finish if the termination condition is met}
\item{Select the gather with the highest disagreement and obtain its picks from the oracle}
\item{Return to step 2}
\end{enumerate}

\section{Results}
\begin{figure}
        \includegraphics{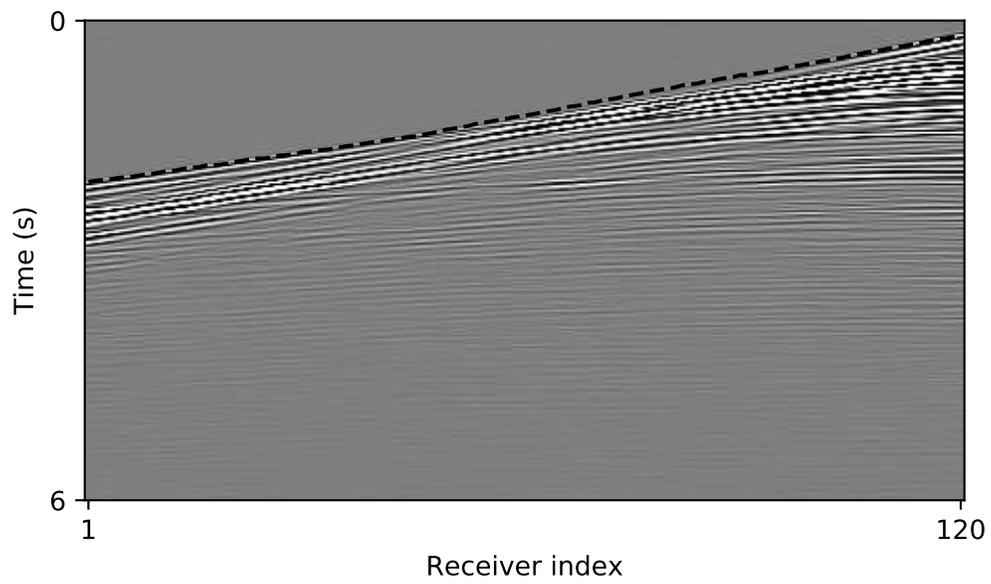}
        \caption{An example shot gather from the test dataset, with the first breaks indicated by a dashed line.}
        \label{fig:shot_gather}
\end{figure}

I test the proposed method by picking first breaks on a 2D marine towed streamer dataset \cite{keys1998comparison}. This dataset contains 1001 shot gathers, each containing 120 traces. Figure \ref{fig:shot_gather} shows an example shot gather. To measure its effectiveness, I compare it with a method that is the same except that the next gather to pick is randomly chosen from the remaining unpicked gathers. To obtain the first break picks that will be considered the true values for this dataset, I use times obtained by manually picking all gathers. For consistency, I use the true picks as the oracle. Each time picks for a gather are requested, the true picks for that gather are thus returned. I perform the picking on one shot gather at a time.

\subsection{Number of gathers considered by the committee}
\begin{figure}
        \includegraphics{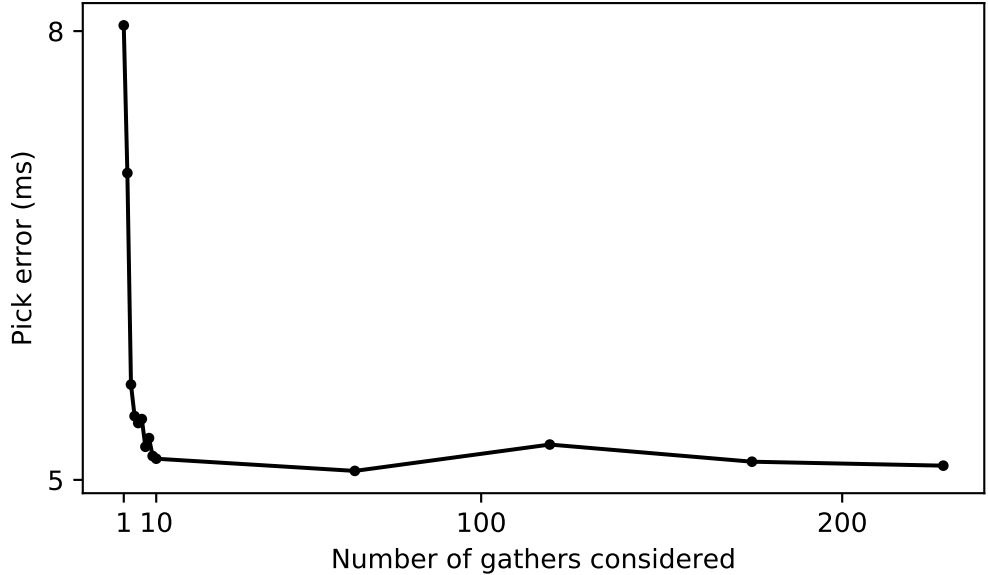}
        \caption{The mean error over the predicted picks for the entire dataset after picking ten gathers decreases with the number of gathers that the committee considers after each gather is picked in order to decide which gather would be optimal to pick next, but the improvement is flat for more than ten gathers. Each trial is repeated thirty times with different random seeds.}
        \label{fig:n_gathers_per_it}
\end{figure}

To determine the effect of changing the number of gathers that are randomly selected to be considered by the committee as possible next gathers to pick, I use the proposed method to pick ten gathers for a range of values of this parameter. I repeat each trial thirty times with different random seeds and calculate the mean absolute error between the predicted and true picks for all traces in the dataset after the tenth gather has been picked. The result, in Figure \ref{fig:n_gathers_per_it}, shows that the error decreases until about 1\% of the dataset is considered, after which the error does not change substantially. This indicates that considering 1\% of the dataset is sufficient, but, for safety, in the remaining results I let the committee consider 10\% of the dataset after each gather is picked.

\subsection{Effect on gather selection}
\begin{figure}
        \includegraphics{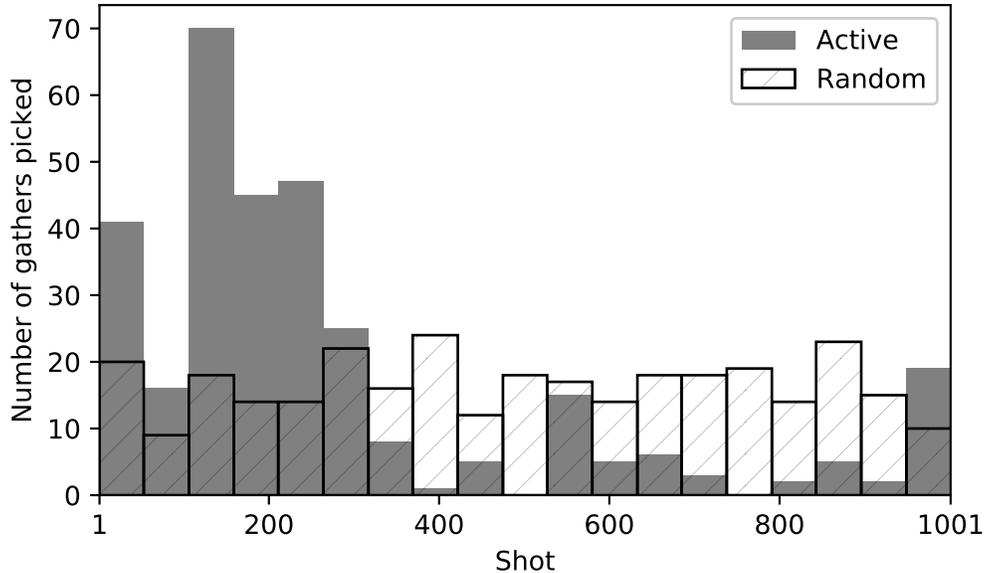}
        \caption{Most gathers selected for picking by the active learning-based method are clustered near shot 200, which is where there is a jump in the first break pick time. The measurements are obtained by repeating a trial fifteen times, where each trial consists of picking twenty gathers, giving a total of 300 gathers selected each for both the active learning and random selection approaches.}
        \label{fig:md_vs_random_queries}
\end{figure}

\begin{figure}
        \includegraphics{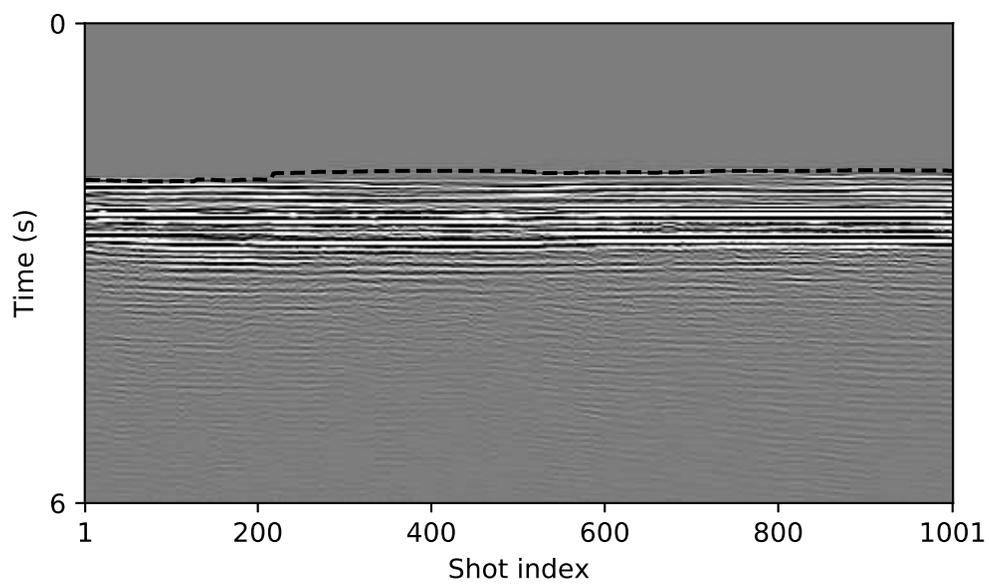}
        \caption{The traces recorded for all shots by one of the receivers, with the first breaks indicated with a dashed line. There is a jump in the first breaks near shot 200 as an earlier arrival becomes visible.}
        \label{fig:channel_gather}
\end{figure}

The goal of using active learning is to optimise which gathers are selected to have their first breaks picked by the oracle. When these gathers are selected randomly, we expect that the selected gathers will be evenly distributed over all gathers in the dataset. We see in Figure \ref{fig:md_vs_random_queries} that this is the case. We also see that the gathers selected by the proposed active learning approach are strongly clustered near shot 200. In Figure \ref{fig:channel_gather} we see that the first break pick time jumps around this shot, as the geology of the area covered by this dataset causes an arrival at an earlier time to be visible beyond it. The position of this jump is uncertain without knowledge of the true picks for the entire dataset, and so it is reasonable for an active learning method to focus on this area.

This could be done by the human operator manually looking through the data in different dimensions to identify the most important gathers to pick, but typical seismic datasets are often too big for this to be feasible. Active learning instead automates the process of finding and prioritising the important gathers to pick.

\subsection{Effect on accuracy}
\begin{figure}
        \includegraphics{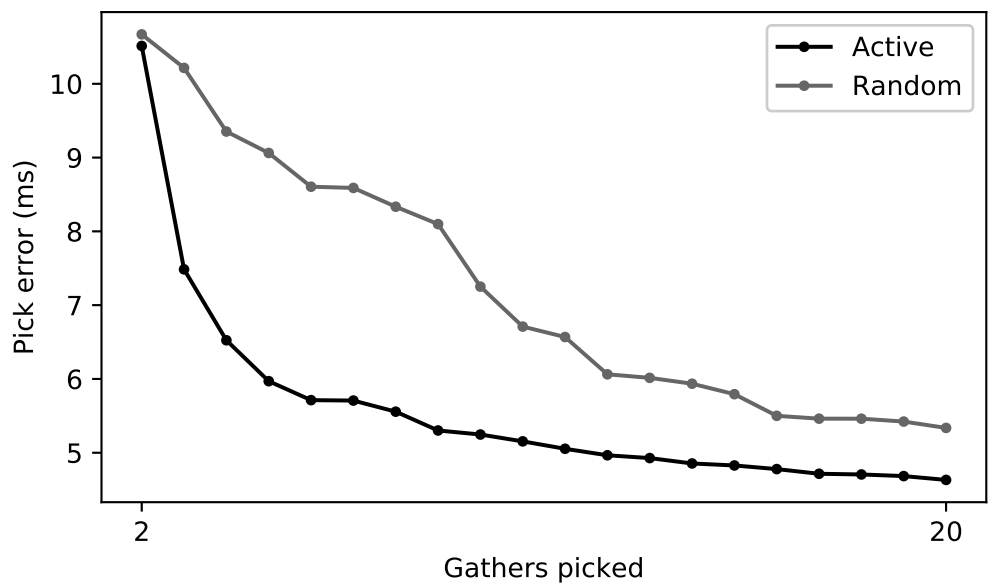}
        \caption{The error in the predicted picks over the entire dataset after each of twenty gathers is picked, shows that the error decreases more rapidly using the active learning-based approach, especially for the early gathers. Each trial is repeated fifteen times and the mean error value is used.}
        \label{fig:md_vs_random_errs}
\end{figure}

Figure \ref{fig:md_vs_random_errs} shows that the error decreases more rapidly when using the proposed approach compared to randomly selecting the next gather.

\section{Conclusion}
Active learning, implemented for the first break picking task using the proposed method, reduces the number of gathers that need to be picked by focusing attention on the regions of the dataset where the values are most uncertain. It achieves this while using understandable algorithms rather than a ``black box'' deep neural network. The method also integrates with existing workflows, replacing picking on a regular grid with a more careful selection of gathers to pick. The computational cost is kept to a moderate level by only selecting a fraction of gathers for consideration by the committee after each gather is picked.

\bibliography{active_pick}{}

\begin{thebibliography}{1}

\bibitem{bekara2019automatic}
Ma{\"\i}za Bekara and Anthony Day.
\newblock Automatic {QC} of denoise processing using a machine learning
  classification.
\newblock {\em First Break}, 37(9):51--58, 2019.

\bibitem{castro2005faster}
Rui Castro, Rebecca Willett, and Robert Nowak.
\newblock Faster rates in regression via active learning.
\newblock In {\em NIPS}, volume~18, pages 179--186, 2005.

\bibitem{keys1998comparison}
Robert~G Keys and Douglas~J Foster.
\newblock {\em Comparison of seismic inversion methods on a single real data
  set}.
\newblock Society of Exploration Geophysicists, 1998.

\bibitem{10.1007/978-3-319-46562-3_24}
Jack O'Neill, Sarah Jane~Delany, and Brian MacNamee.
\newblock Model-free and model-based active learning for regression.
\newblock In Plamen Angelov, Alexander Gegov, Chrisina Jayne, and Qiang Shen,
  editors, {\em Advances in Computational Intelligence Systems}, pages
  375--386, Cham, 2017. Springer International Publishing.

\bibitem{sabbione2010automatic}
Juan~I Sabbione and Danilo Velis.
\newblock Automatic first-breaks picking: New strategies and algorithms.
\newblock {\em Geophysics}, 75(4):V67--V76, 2010.

\bibitem{settles2012active}
Burr Settles.
\newblock Active learning.
\newblock {\em Synthesis lectures on artificial intelligence and machine
  learning}, 6(1):1--114, 2012.

\end{thebibliography}
\bibliographystyle{plain}
\end{document}